\documentclass[12pt]{article}
\usepackage{bm}
\usepackage{amsmath}
\usepackage{amssymb}
\usepackage{graphicx}
\usepackage{amsmath}
\usepackage{bbm}

\begin{document}

\baselineskip 15pt

\title{\bf Does quantum nonlocality irremediably conflict with Special Relativity? }

\author{GianCarlo Ghirardi\footnote{e-mail: ghirardi@ts.infn.it}\\ {\small
Department of Theoretical Physics of the University of Trieste,}\\ {\small the Abdus
Salam International Centre for Theoretical Physics, Trieste, and}\\{\small  Istituto Nazionale di Fisica Nucleare, Sezione di Trieste, Italy.}  }

\date{}

\maketitle

\begin{abstract}
We reconsider the problem of the compatibility of quantum nonlocality  and the requests for a relativistically invariant theoretical scheme.We begin by discussing a recent important paper by T. Norsen on this problem and  we  enlarge our considerations  to give a  general picture of the conceptually relevant issue to which this paper is devoted. 

\end{abstract}

\vspace{2cm}

\section{General considerations}
The celebrated paper by John Bell \cite{bell} which, combined with many experimental investigations, has made evident that nature exhibits some nonlocal features has compelled the scientific community to reconsider the problem of the compatibility of the two \cite{bell1} {\it fundamental pillars of contemporary theory}, quantum mechanics and relativity. The situation is worsened by the fact that serious misinterpretations of the real significance of Bell's work characterize the debate on this problem and that many, even brilliant, physicists have never grasped various subtle points of Bell's analysis.

Fortunately, recently some important and clear-cut contributions to this problem have appeared, a significant example among them being  a paper by T. Norsen \cite{norsen} which has stimulated us to perform the analysis we are going to present here.  The paper focusses with remarkable lucidity the problem of Q-nonlocality and it calls the due attention to many subtle points of Bell's analysis with particular reference to  the position taken by Jarrett \cite{jarrett} on this problem.

 Here, we take a slightly different perspective, we  try to clarify  some aspects which deserve to be further deepened and, finally, we   discuss some questions which have not been considered in ref.[3]. In particular,  Norsen, having as his primary objective the comparison of Bell's position with the one taken by Jarrett, has not mentioned various statements by Bell on the problem of the compatibility of  nonlocality with the requirements of  Special Relativity (SR). Correspondingly, he  has concentrated his attention mainly to the precise position taken by Bell in one of his last papers \cite{bell2} without mentioning others  significant writings by him which deal with   the question we are interested in. The most serious limitation of this choice, in our opinion, is that it does not allow to keep clearly distinct the logical status of different locality requirements and, above all, to fully appreciate Bell's position concerning the relativistic generalizations of what he considered \cite{against} {\it  the only two exact pictures} of quantum processes and, consequently, of grasping the crucial points of the present debate on the problem.

The paper is organized as follows. In the next Section we   list a set of points which, in spite of the fact that they are often ignored in debates on the subject, we consider as definitively established. They have been exhaustively discussed in ref.\cite{norsen} and  we do not intend to analyze them further. Section 3 is devoted to outline the  general formal schemes which we will consider in this paper, schemes  which   satisfy all the conditions which Norsen has so appropriately identified as essential requirements  put forward by Bell for any theoretical account of natural phenomena.  In the same section, by resorting to the mathematical notation which we will utilize, we  recall Bell's locality condition  and its ``decomposition" into two logically independent assumptions which is at the basis of Jarrett's analysis and which will be essential for the subsequent discussion. We also present some conclusions concerning theories which violate one or the other of the just mentioned locality conditions. Section 4  will summarize Jarrett's position  and the appropriate criticisms that Norsen has put forward concerning his conclusions. Starting from this point   our position will  depart from the one of ref.\cite{norsen}.  The  subsequent section  deals with a reconsideration of Bell's position concerning relativistic requirements and  makes clear our point of view  by making  reference to  precise statements by John Bell. In the last Section some conclusive remarks are presented.

We hope that the present analysis, supplementing the one  of ref.[3] to which we will make continuous reference, will contribute to a further clarification of the many subtle points related to the problem under examination.

\section {Basic  implications of Norsen's analysis}

There are various relevant points which have been  lucidly discussed in ref.[3] and/or have been focussed in the writings of John Bell. We consider it appropriate to list them here for making clear what we consider firmly established and, as such, not deserving  further  consideration.
\begin{itemize}
\item A common misinterpretation of Bell's work derives from the (mistaken) idea that the derivation of his celebrated theorem relies not just on the requirement of locality but also on some other assumptions, such as  Realism, Hidden Variables, Determinism or Counter-Factual-Definiteness. We consider quite frustrating that even extremely brilliant  scientists seem to  take such an incorrect position \cite{zeilinger,aspect} in particular including the assumption of realism among Bell's hypotheses.
\item The analysis of Bell's theorem requires the consideration of precise experiments on appropriate systems. Such experiments are characterized by the settings of the  apparata used to perform the measurements. It is a fundamental assumption of Bell's derivation that these settings \cite{bell2} {\it  can be considered to be free, or random}. This is often related to the fact that, in the relevant type of experiments, there are two far away experimental physicists who can freely choose ``at the last minute" such settings\footnote{Obviously, as remarked by Bell himself \cite{bell2} {\it one can envisage theories in which there just are no free variables ... In such ``superdeterministic" theories the apparent free will of experimenters, and any other apparent randomness would be illusory.} A paradigmatic case of this type would be the assumption that a ``pre-established harmony" characterizes all physical events. If it is codified in the initial condition of the universe that John Bell will derive his theorem and that Alain Aspect will verify it at a precise time, following a precise protocol and obtaining the precise results he did  obtain, etc., then the whole argument breaks down, and the scientific enterprise  looses any meaning, even though it would be impossible not to pursue it, since also this search would be unavoidably determined by the initial conditions.}. 
\item We will deal systematically with local experiments ``which have  outcomes". We are perfectly aware of the remarkable difficuties that  this problem meets within standard quantum mechanics. We also know, and we will devote a part of our work to analyze them, that there are consistent ways \cite{bohm,grw} of circumventing it.  Here we do not want to committ ourselves to this delicate point. For our purposes what matters is simply that, according to our experience and in the situations we we will consider, local ``measurement" procedures actually lead to definite outcomes and that the probabilities of their occurrence are those predicted by the standard theory with the (inconsistent) postulate of wave packet reduction\footnote{Here, the statement that definite outcomes are known to emerge, makes reference to the fact that, at the end, all those processes which we denote as measurements are  related in a fundamental way to perceptions. We do not want to be misunderstood: we do not intend to attribute any peculiar role to human observers. We take a position like the one taken by Bell \cite{bellGRW} when he has presented his first suggestion concerning the beables of the GRW theory. He stated: {\it the GRW jumps are well localized in ordinary space. So we can propose these events as the basis of the `local beables' of the theory. These are the mathematical counterparts in the theory to real events at definite places and times in the real world. A piece of matter then is a galaxy of such events. As a schematic psychophysical parallelism we can suppose that our personal experience is more or less directly of events in particular pieces of matter, our brains, which events are in turn correlated with events in our bodies as a whole, and they in turn with events in the outer world.} If one accepts, as we do, this clear position,  when  speaking of outcomes one is always making reference to perceptually definite situations.   Now, there is no doubt that if one has an ``apparatus" working with 100\% efficiency and one triggers it (using for simplicity the standard quantum language) with one of the eigenstates of the operator associated to the variable one wants to study, he ends up with the perception corresponding to the registration of the corresponding eigenvalue. In a similar way,  if one triggers the apparatus with a superposition of such eigenstates one still ends up with a perception corresponding to one of the possible eigenvalues, with the probabilities given by quantum mechanics. These remarks should make clear why, in the first part of our analysis which deals exclusively with (individual and correlated) outcomes, we will avoid  committing ourselves too strictly to the fundamental problem of describing how the collapse can occur.}. 
\item All our considerations will be restricted to theories in which space-time can be regarded as given and in which it is meaningful to speak of localized events. The reason for this is that, just to mention some examples, when space-time  is ``quantized" or when one adopts the  position characterizing the presently very fashonable ``string theories of everything", the concept of locality itself becomes very obscure. Consequently it turns out to be impossible to attach any precise meaning to the very problem we are interested in.
\item In our analysis  we  make  use of  terms such as Locality, Causality, Outcomes, States,  etc.,  which make reference to  fundamental aspects of our description of the world. It is a great merit of ref.[3] to have called the attention of the reader on the fact  that  these crucial expressions are meaningful only with reference to some candidate theory. We hope to succeed in making this clear during our discussion by stressing the specific logical status of the terms we will use in the various contexts of our discourse.
\item The previous remark does not diminuish in any way whatsoever the extreme relevance of Bell's theorem. What he has shown is that any physical theory satisfying an extremely general request of Locality makes predictions concerning a certain class of experiments which, being constrained to respect Bell's inequality, violate the predictions of Quantum Mechanics. This statement seems to conflict, to some extent, with the asserted dependence of the whole conceptual picture on the structure of the specific theory one is considering. And here one cannot avoid recalling the extremely appropriate and lucid remarks by T. Norsen:
\begin{quote}
{\it  How then did Bell think we  could end up saying something interesting about Nature? That is precisely the beauty of Bell's theorem, which shows that no theory respecting the locality condition can agree with the empirically verified QM predictions for certain types of experiments. That is, no locally causal theory in Bell's sense can agree with experiments, can be empirically viable, can be true. Which means the true theory (whatever it might be) necessarily violates Bell's locality condition. Nature is not locally causal.}
\end{quote}

\end{itemize}

\section{The formal  framework and the locality requirements}
\subsection{General aspects}

Let us begin by defining in a precise way the class of theories we take into consideration as potential candidates for the description of natural phenomena. Since Galileo's times we have learned that it turns out to be particularly appropriate  resorting to the use of mathematical entities in order to
 describe precise  situations of the physical systems we are interested in. Having assumed an underlying space-time structure we will consider our system at a given time (a $t=${\it const} suface or, more generally, a space-like surface) and we will characterize its ``state" in terms of two types of mathematical entities which we  denote with the letters $\mu$ and $\lambda$, and which we  call ``controllable" and ``uncontrollable" variables, respectively.
 
 Some specifications:
 \begin{itemize}
 \item We have resorted to the expressions ``controllable" and ``uncontrollable" to indicate that, in principle, we are able, by resorting to appropriate physical procedures, to prepare the system in such a way that its controllable variables take a precise value among the possible ones, while we have no precise control of the uncontrollable variables.
 \item We assume  that, even though we  have not  a full control of the unaccessible variables, we  have some knowledge of the way they turn out to be distributed as a consequence of our ``preparation procedures".
 \item We assume, and this is a crucial point of our analysis, that the assignement of the variables $\mu$ and $\lambda$ represents {\bf the most accurate specification} which,  within the considered formal scheme, is {\bf in principle} possible for the state of an individual physical system\footnote{To make our choice clear to everybody we recall that Standard Quantum Mechanics identifies $\mu$ with the statevector $\psi$ and assumes that there are no uncontrollable variables, while Bohmian mechanics  assumes once more that $\mu=\psi$ while it identifies the uncontrollable variables $\lambda$ with the positions of the particles of the system.}. We note that this perspective makes clear, as already stressed, that the concept of state turns out to be fundamentally dependent on the specific candidate theory we consider.

We believe it might be useful to illustrate the idea of {\it the most accurate specification} used above by considering the well known example  of two marbles of different colours  lying in two boxes. The  boxes are  very far from each other. Would one assume that all what is known about the state of the system is that there is one black marble and one white marble and that the boxes are  A and B, respectively, one would be led to conclude that the probabilities (with obvious meaning of the symbols) $P(Black,A)$ and $P(White,B)$ are both equal to $1/2$. Since, however, the joint probability $P([Black,A] \& [White,B])$ also equals $1/2$, it does not satisfy the {\it factorizability} request\footnote{In this Section we will make use of some technical expressions like ``factorizability" and ``completeness" which are well known and which we will make precise in what follows.} (which is strictly related to locality, see below), i.e. that the probability that ``Alice, opening the box  will get the outcome: {\it The marble in box A is black}, {\bf and} Bob the one: {\it The marble in box B is white}", is  the product of the probabilities of the  two considered instances. 

Some comments are appropriate:

\begin{itemize}
\item It goes without saying that if our theory deals with a universe  in which all the marbles have definite colors and locations and they are accessible variables, then the most accurate specification of the state of our physical system would not be the one given above but, e.g.,  "There is a Black marble in A and a White one in B" and in such a case the  probability of this joint outcome would be  1 while the one of the outcome corresponding to an exchange of the colors associated to the boxes would be 0. In both cases the {\it factorizability} condition would be satisfied since the single outcome probabilities would take both the value 1 in the first case and both the value zero in the second. Accordingly, in the considered case no violation of locality occurs and we can disregard it. Before proceeding, we simply call attention to the fact that  the just mentioned violation of  {\it factorizability}  is entirely due to the fact that the specification of the state  is not ``the most accurate" one which is possible in principle, i.e., it is due to the {\it Incompleteness} of the  description (see next Section). In this case {\it factorizability} would not hold, but no nonlocal effect takes place, in spite of the fact that the knowledge of one outcome  changes the probability assigned to the other one. 
\item One might also imagine that the hypothetical theory he is considering refers to a situation (a universe) in which there is no way, in principle, to know the ``color-box" association because it is, e.g., chosen by some sort of divinity or it is made unaccessible to us by the intrinsically ``veiled" nature of reality \cite{d'espagnat}. In this scenario, in which  the marbles have definite (even though unknowable) locations, we have a more basic violation  of {\it factorizability} which, however, derives  again from an incomplete, even though unavoidably so,  knowledge of the state of the system.
\item It is with reference to  a situation strictly similar to those just mentioned which Jarrett concludes that a violation of {\it completeness} does not conflict with SR. We fully agree with Maudlin \cite{maudlin} and Norsen that his argument is not compelling and that  it is incorrect to make reference, as Jarrett does, to  cases  with the indicated features to conclude that, in general, violating {\it completeness},  does not imply a conflict with SR.
\end{itemize}

\end{itemize}

\subsection{The locality issue within our scheme}

With these premises we can now tackle explicitly our problem. We will be interested in a situation in which two far away systems are subjected to local experiments aimed to ascertain the value of some physical variable, i.e., one of those quantities which are usually referred to as ``observables"\footnote{We are perfectly aware that John Bell definitively preferred to use the term ``beable" in place of the ambiguous term ``observable", and we agree with him concerning the unsatisfactory status of the last expression. Moreover, as we will make clear in what follows, we completely share his position as well as the one of ref.\cite{allori} that   the specification of its ``primitive ontology", amounting to the identification of ``what the theory is fundamentally about", or, in Bell's language, of which quantities are posited as beables, is an essential part of any theoretical scheme. 

However, both with reference to writings of Bell \cite{bell2,bell3} as well as to the use made by Norsen \cite{norsen}, we would like to point out that it seems to us not completely satisfactory to resort to the expression beable to characterize, among other things, also what we have denoted as the state of the system. In fact in clarifying the meaning he was attaching to the term beable Bell has  stated that, in his opinion: {\it Indeed, observations and observers must be made out of beables}. Since in what follows we will consider candidate theories like quantum mechanics or dynamical reduction models in which the states of physical systems are specified by  wavefunctions, it seems unappropriate to us to claim that ``observers are made of wavefunctions". For these reasons we prefer to use the ``imprecise" term ``value of a physical variable" to denote the outcomes.}.  Different experiments will be specified by different ``settings", and, following ref.\cite{bell2}, we will denote the settings of the detectors in the two space-like regions 1 and 2 in which the experimental tests are performed by the symbols $a$ and $b$, respectively. Correspondingly, we will denote as $A$ and $B$ the outcomes corresponding to the tests which are actually performed. The situation is illustrated in Fig.1.

\begin{figure}[h]
\begin{center}
\includegraphics[scale=0.6]{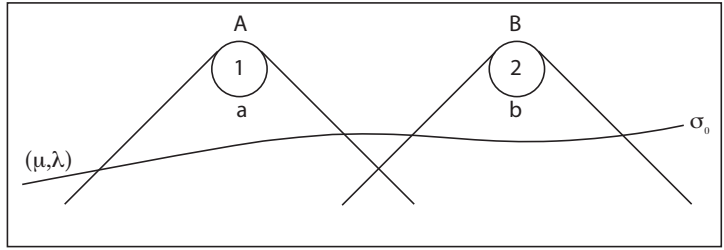}

\caption{The typical physical situation to which our arguments will make reference. The meanings of the various labels are explained in the text.}  \label{f1}
\end{center}
\end{figure}

Before going on we need to make precise an essential element of  Bell's argument which is often underappreciated and to which Norsen has appropriately devoted a great attention, i.e. the fact that the most accurate  specification of the state of the system must be such that it shields off  each of the two space-like regions 1 and 2 from the overlapping of the backward light cones from such regions. We refer the reader to ref.\cite{norsen} for a detailed discussion of this crucial point. For our purposes one can read it in its simplified form as stating that the $t=const$ or  space-like surface $\sigma_{0}$ on which the ``conceivably most accurate specification of the state of the system is given" does not intersect the region  common to the two cones, as shown in the Figure.

Now comes the fundamental assumption of our analysis. We assume that the assignement of the variables $\mu$ and $\lambda$, i.e. the specification of the state of the system on the ``initial" space-like surface and the dynamics of the theory we are dealing with, determines all single and joint, conditional and unconditional probabilities of the outcomes of the experimental tests which can be performed in regions 1 and 2. We can then consider \cite{bell2} the joint probability $P(AB\vert a,b;\mu,\lambda)$ and express it through the conditional probability   $P(A\vert a,b;B;\mu,\lambda)$ as follows\footnote{Obviously the role of $A$ and $B$ in Eq.(1) can be interchanged.}:
\begin{equation}
P(AB\vert a,b;\mu,\lambda)=P(A\vert a,b;B;\mu,\lambda)P(B\vert a,b;\mu,\lambda)
\end{equation}
At this point we proceed with specific requests of locality. First of all we impose the condition which has been denoted as {\it Completeness} by Jarrett \cite{jarrett} and {\it Outcome Independence},\{O.I.\},  by Shimony \cite{shimony}:
\begin{eqnarray}
P(A\vert a,b;B;\mu,\lambda) & = & P(A\vert a,b;\mu,\lambda),\nonumber \\
P(B\vert a,b;A;\mu,\lambda) & = & P(B\vert a,b;\mu,\lambda),
\end{eqnarray}
and secondly we impose {\it Locality} (Jarrett) ({\it Parameter Independence}, \{P.I.\},  (Shimony)):

\begin{eqnarray}
P(A\vert a,b;\mu,\lambda) & = & P(A\vert a;\mu,\lambda)\nonumber \\
P(B\vert a,b;\mu,\lambda) & = & P(B\vert b;\mu,\lambda).
\end{eqnarray}

Relation (1) and conditions\footnote{From now on, for reason of simplicity, we will use the symbols \{O.I.\} and \{P.I.\} to denote Jarrett's Completeness and Locality, respectively.} (2,3) lead then to the {\it Factorizability} condition:
\begin{equation}
P(AB\vert a,b;\mu,\lambda)= P(A\vert a;\mu,\lambda)\cdot P(B\vert b;\mu,\lambda).
\end{equation}

As  already stated,  the uncontrollable variables $\lambda$ have a certain probability distribution, which we  denote as $\rho(\lambda)$. The requirement that the settings $a$ and $b$ of the apparata can be chosen at free will by the experimenters\footnote{Actually, the only relevant feature for the  argument is that the settings are not influenced by the variables characterizing the state.} enters into play to allow us to relate the probabilistic predictions of our candidate theory to those which characterize actual experiments, which in turn we assume to coincide with those implied by standard quantum mechanics.  Let us define  a correlation function $E(a,b,;\mu)$ as the expectation value of the product\footnote{In the following equation we assume that the unaccessible variables $\lambda$ are continuously distributed over a set $\Lambda$. Obviously nothing changes in the  discrete case.} of $A$ and $B$:
\begin{equation}
E(a,b;\mu)=\sum_{A,B}\int_{\Lambda} A\cdot B \cdot P(AB\vert a,b;\mu,\lambda)\rho(\lambda)d\lambda.
\end{equation}

Let us denote as $\{{\it Fact}\}$  assumption (4). Then one sees from Eq.(5) that  $\{{\it Fact}\}$ implies:\begin{equation}
E(a,b;\mu)=\sum_{A,B}\int_{\Lambda}A\cdot B \cdot P(A\vert a,\mu,\lambda)P(B\vert b,\mu,\lambda)\rho(\lambda)d\lambda,
\end{equation}
\noindent a relation which leads in a straightforward way to the CHSH \cite{clauser} inequality:
\begin{equation}
\vert E(a,b;\mu)-E(a,b';\mu)\vert+\vert E(a',b;\mu)+E(a',b';\mu)\vert\leq 2.
\end{equation}

It goes without saying that one has to identify the correlation function of Eq.(5) for a given value of $\mu$ (which is the analogous of the controllable  statevector of Q.M.) with the quantum mechanical expectation value $\langle\mathcal {A} \mathcal{B}\rangle_{\psi}$ 
 of the product of the two self-adjoint operators $\mathcal {A}$ and $\mathcal{B}$ associated to  the test procedures (characterized by the settings $a$ and $b$) of the considered experiment when the quantum state of the composite system is $\psi$. But, according to quantum mechanics, the analogous of the inequality (7) for the expectation values is violated for appropriate choices of the settings $a,a',b,b'$ when the state $\psi$ of the two far away systems is entangled.

Concluding: the quantum predictions for the outcomes of appropriate correlation experiments cannot be accounted for by any member whatsoever of the set of candidate theories satisfying our locality requests.

Some remarks are at order.
\begin{itemize}
\item The class of candidate theories we have considered is extremely large. The basic assumptions are that such theories contain either accessible or inaccessible (or both) types of variables giving the most accurate (in principle) description that the theoretical framework considers as characterizing a given physical situation, that the experimenters can freely choose the settings of their apparata and that these settings are not influenced by the  just mentioned variables. One also assumes that the state of the system is specified  in such a way that the two space-time regions which are involved are screened off from each other by the assignement of the initial situation.
\item The really crucial assumption which renders the scheme unviable for accounting of natural processes (if they are assumed to be governed by the laws of quantum mechanics) is\footnote{One might remark that $\{{\it Fact}\}$ does not fully grasp the idea of Locality, because it reduces essentially to the request of no-correlations between space-like events. Actually, such a remark has been made to the present author by V. Tausk, who has called attention that even though $\{{\it Fact}\}$ holds, there might be underlying nonlocal interactions which do not violate such a request. I believe that this criticism is correct but not pertinent; actually it gives even more strength to the conclusion in the sense that it enlarges the class of theories incompatible with quantum mechanics. Also theories which would allow instantaneous underlying interactions, provided they do not spoil $\{{\it Fact}\}$,  cannot agree with quantum mechanics. Even on physical grounds it seems to us that, for the analysis we are interested in,  one can safely ignore such peculiar interactions provided they do not induce correlations. We would also like to stress that since from the identity (1) one has that $\{O.I.\}\wedge \{P.I.\}\Rightarrow\{{\it Fact}\}$ one also has that $\neg\{{\it Fact}\}\Rightarrow \neg \{O.I.\}\vee\neg\{P.I.\}$, i.e., if  $\{Fact\}$ does not hold,  then at least one of the two locality requests must be violated.} condition  (4).
\end{itemize}

\subsection{General remarks concerning \{O.I.\} and \{P.I.\}}
Let us suppose that $\{Fact\}$ is violated. Then, as we have seen, either \{O.I.\} or \{P.I.\} or both must be violated. The first crucial remark derives from considering the whole class of nonlocal (in the sense that they violate the factorizability request) deterministic theories predictively equivalent to QM. Since no deterministic theory agreeing with quantum mechanics can violate \{O.I.\}  (just because $0\cdot 0=0\cdot 1=1\cdot 0=0,1\cdot 1=1$) if it violates $\{Fact\}$ then it must violate \{P.I.\}.

It is easy to prove \cite{ggrassi} that, in order to generalize a theory violating \{P.I.\} in such a way that it respects the basic  relativistic requirement that all observers are on the same footing, one must resort to a preferred (hidden) reference frame.  The same does not hold, in principle, for a theory which violates only\footnote{We are  not claiming that  a theory which violates only  \{O.I.\} cannot be characterized by the existence of a preferred reference frame; actually, it would be easy to build such a theory. We claim that one cannot prove that any relativistic generalization of a nonrelativistic theory violating \{O.I.\} requires to resort to  a preferred frame.} \{O.I.\}.

We consider important in order to understand our subsequent analysis to keep clearly in mind the just mentioned conceptual difference between the status of the two locality requests whose conjunction implies $\{Fact\}$.

\section{Nonlocality vs Special Relativity, first conclusions}
We will consider now the debate concerning the incompatibility or the peaceful coexistence of the  nonlocal aspects of natural processes with the requirements imposed by Special Relativity, and we will summarize Jarrett's and Norsen positions about it.

\subsection{Jarrett's claims}
The central, and most important part of Norsen's paper is devoted to confront Bell's and Jarrett's positions concerning the implications of nonlocality for a relativistic description of natural processes. Let us begin by summarizing what Norsen takes as Jarrett's position.
\begin{itemize}
\item
Jarrett, investigating the status of Bell's locality requirement with respect to the constraints imposed by SR, has reached the conclusion that Bell's condition is, in a precise sense, too strong. He has remarked, as we have shown above, that such a condition corresponds to the joint requirements of \{O.I.\}, Eqs.(2) and  \{P.I.\}, Eqs.(3).
\item According to Jarrett, a violation of \{P.I.\} implies the possibility of faster than light communication or instantaneous action at a distance, which is obviously  incompatible with SR.
\item On the contrary, a violation of \{O.I.\} would not conflict with relativity. Jarrett, to enforce this conclusion discusses a situation  analogous to the one of the two marbles and the two boxes  in the case of an incomplete specification of the state (in which box is the ball of a given color). For all those who have followed our arguments it will be obvious that if one specifies only  the colors of the balls and the location of the boxes, one has a violation of \{O.I.\}  due to {\it incompleteness}, but, if one assumes that, prior to any test, each marble is in a precise box, no conflict arises between SR and the fact that the probability of a far away outcome changes instantaneously as a consequence of the knowledge of the outcome at the other wing of the apparatus. This is the limited and inappropriate  reading that Jarrett gives of the situation in which a theory violates $\{Fact\}$ just because it violates \{O.I.\}.
\end{itemize}

\subsection{Critics to Jarrett's position}
Norsen correctly calls attention to various elementary  facts which show the inappropriateness of the points made by Jarrett. Let us mention them.
\begin{itemize}
\item In order that a violation of \{P.I.\} leads to faster than light effects one must at least have access to the uncontrollable variables of the theory. The paradigmatic example is represented by Bohmian mechanics. In it, faster than light signaling is possible only if one has access to the actual position of the particles. If this is not the case the unfolding of any physical process leads to the same outcomes as standard QM in which, as well known \cite{eberhard,GRWFTL}, one cannot take advantage of nonlocality to induce testable instantaneous changes at-a-distance.
\item The ``proof" that a violation of \{O.I.\}  does not lead to a conflict with SR presented by Jarrett rests strictly on the absolutely elementary and inappropriate case he has considered. As it is obvious, and as remarked also by Maudlin\cite{maudlin}, one can easily imagine cases in which one local event which leads to an intrinsically random outcome is accompanied by the emission of a tachion which reaches the far away partner system and induces the observed correlations. According to Norsen, this way of violating \{O.I.\}   in no way eliminates the fact that the theory  violates relativistic requirements (even in the case in which it would not allow an actual  faster-than-light signaling) because of the occurrence of instantaneous causation at a distance. This remark is basically correct, but it deserves a more detailed analysis.
\item The most relevant criticism to Jarrett's position is that, while it is true that Faster than Light effects conflict with SR, it is totally inappropriate to reduce the requirements of SR to the one that the theory under discussion does not allow such effects.
To analyze this important point Norsen embarques himself in a detailed analysis of Bell's precise position concerning what requirements should be imposed to a theory in order to be allowed to consider it not conflicting with SR. Now, while we fully agree that the problem of the compatibility with SR cannot be reduced to the impossibility of superluminal signals, we believe that Norsen has not been sufficiently general in describing Bell's position. More about this in what follows.
\item Norsen claims that, in any case, a violation of \{O.I.\} enters into an irremediable conflict with SR, provided the other specifications made by Bell concerning the total experimental situation are taken into account\footnote{We consider  it appropriate to stress that in introducing our general formal scheme we have guaranteed from the very beginning that we will limit our considerations to candidate theories satisfying all further requests (screening off, free will etc.) advanced by Bell.}. While we fully agree with Norsen that Jarrett has been too superficial in his analysis drawing  unjustified conclusions, we do not agree  with his point of view concerning the status of \{O.I.\}. In particular we consider not justified  his choice to put on the same footing the violation of  \{O.I.\} and the one of \{P.I.\}.
\end{itemize}

\section{A more general analysis}

In this  section we intend to reconsider the whole problem of the relations between nonlocality and SR by making  specific reference to Bell's general views regarding the conditions which make  a theory  compatible with relativistic requirements. To this purpose we consider it useful to enrich Norsen's presentation by mentioning Bell's statements concerning a possible position about relativity which he has clearly identified as logically consistent even though he did not  share it. This discussion should be relevant both to fully appreciate the modern positions with respect to relativistic theories which overcome the macro-objectification problem as well as to  prepare the grounds for putting forward our criticism concerning Norsen's conclusions about the implications of \{O.I.\}.

\subsection{Parameter Dependence and relativistic requirements}
We begin by remarking that in exposing Bell's views on the matter,  Norsen has made reference almost exclusively to the position taken by this great thinker in one of his last papers: {\it La Nouvelle Cuisine} \cite{bell2}. We think that it is the choice of considering this paper as the ``bible"  concerning Bell's views on the relations between nonlocality and SR which has led Norsen to put forward statements as:
\begin{quote}
{\it Bell's own interpretation of the meaning of his theorem ... undermines the attempt to establish the Peaceful Coexistence of QM and SR} 
\end{quote} 
 or  to invoke a sentence of ref.\cite{bell1} to assert that according to Bell
 \begin{quote}
{\it  to postulate a dynamically privileged reference frame} would represent {\it ``a gross violation of relativistic causality".}
\end{quote} 

With reference to these claims, we consider it appropriate to  recall that while Bell has certainly expressed his clear preference for what he has denoted as  ``genuine invariance"  (an expression which corresponds to the position attributed to him by Norsen), he has never ignored that alternative positions with respect to the relativistic requirements can legitimately be taken. 

For instance it seems particularly relevant to recall that in \cite{schwinger}, in commenting on possible generalizations of Bohmian mechanics, he has claimed:
\begin{quote}
{\it   The} [Bohmian] {\it  picture can be developed for the $\varphi$'s} [quantum fields]  {\it just as for the $q$'s} [positions]. {\it  It accounts then for the phenomena accounted for by relativistic quantum field theory. But it is not Lorentz invariant, in the sense that it has in its structure a preferred frame of reference. If you made the same construction in some other Lorentz frame, you would get different trajectories. However, the theory agrees with all the experiments that agree with relativistic quantum field theory - in particular, it will predict a null result for the Michelson-Morley experiment. So you have something like the situation in relativity theory before Einstein, where there was a preferred frame of reference - there was a significance in absolute simultaneity - but the Fitzgerald contraction and the Larmor dilation fooled moving observers into thinking that light had the same velocity relative to them - so that they could even imagine themselves to be at rest. Now for me, this is an incredible position to take - I think it is quite logically consistent, but when one sees the power of the hypothesis of Lorentz invariance in modern physics, I think you can't believe in it. ... However, what I want to insist is that this theory agrees with experiment - when you dismiss this theory on the grounds of non-Lorentz invariance, you are requiring more than agreement with experiment. }  
\end{quote}

This passage makes quite clear the position of Bell (which to some extent mirrors the one attributed to him by Norsen) concerning SR but it also puts into evidence  that he did not  consider {\it a gross violation of relativistic causality}  the introduction of a preferred reference frame. Actually, the above sentence makes precise that he would not have rejected {\it a priori} a theory exhibiting {\it Parameter Dependence} provided it would  not offer any mean to put into evidence the observer's state of motion. No doubt that ``simplicity" or ``elegance" requirements as well as the adoption of the ``modern" perspective about SR suggest to look for a ``genuinely invariant" theory, but, in his clearly expressed opinion, one surely can have {\bf a quite logically consistent} theory exhibiting {\it Parameter Dependence}. 

Our remarks on this point do not aim to convince the reader that a violation of \{P.I.\} can be easily accepted but to call his attention on the fact that ignoring this lucid statement by Bell might lead to ignore or even to dismiss the relativistic generalizations of Bohm's like theories, a quite relevant aspect of the modern researches on foundational issues. 

We pass now to discuss the, in our opinion, most problematic point of Norsen's analysis, i.e., his position concerning  theories which do not exhibit {\it Parameter Dependence} but only {\it Outcome Dependence}.

\subsection{Outcome Dependence and relativistic requirements}
As already remarked, according  to Norsen  the two conditions of \{O.I.\} and \{P.I.\} have a quite similar logical status. This position is stressed \cite{norsen} in  the following sentence:
\begin{quote}
{\it The most that could be said to distinguish the two subconditions is that, since b} [the setting in region 2 ] {\it is (by definition) controllable and B} [the outcome in region 2] {\it (most likely) isn't, a violation of \{P.I.\} is (all other things being equal) more likely to yield the possibility of superluminal signaling than a violation of \{O.I.\}. } 
\end{quote}
%------------------------------------------------------------------------------------

 \noindent To be fair, we should mention that Norsen himself has felt the necessity to add: \begin{quote}{\it But that only matters if we drop what Bell calls ``fundamental relativity" and instead read SR instrumentally, as prohibiting superluminal signaling but allowing in principle superluminal causation (so long as it can't be harnessed by humans to transmit messages). That is, at best, a dubious and controversial reading of SR, as already mentioned}.
 \end{quote}
 
 However, this specification is not quite appropriate for dealing in general with our basic question. The real and crucial issue under discussion is whether it is possible to work out theories which, even when the most accurate (in principle, not in practice) specification of the state is given,  violate only \{O.I.\} and  do not require the introduction of a preferred reference frame. Taking such a perspective, we believe that Jarrett's splitting of $\{Fact\}$ into the two requests we have discussed has a great conceptual relevance, just because these requirements have a different logical status for the implications concerning a relativistic position about natural processes. 
 
 It seems to us that, in the light of what we have just discussed,  the analysis of ref.\cite{norsen} following the above quoted statement  is not sufficiently articulated. Few paragraphs after this passage, to further support and illustrate the above claim concerning  the status of the \{O.I.\} requirement, the author resorts to the consideration of the toy model discussed by Maudlin \cite{maudlin} which we have already  mentioned. Now, while, as already stressed, we fully agree that  Jarrett's choice of reading the violation of \{O.I.\} in terms of an {\it Incomplete} information concerning the state of a system is inappropriate and inconclusive, we also believe that to criticize it by making reference to a specific and peculiar example  is not particularly meaningful\footnote{All of us are aware of the fact that, given a certain theory which does not exhibits instantaneous effects at a distance, one can devise physically equivalent theories contemplating such effects.}. 

There is also another aspect which we consider particularly relevant. With  reference to the appropriate choice made by Norsen of  resorting to the illuminating comments of such a great scientist as John Bell, we stress that  he has left unanalyzed an extremely relevant point which effectively illustrates the deepness  of Bell's understanding of the conceptual status, when one takes a relativistic perspective, of nonlocal theories which violate only \{O.I.\}. Assuming that the violation of $\{Fact\}$ arising from the violation of the one or the other of the two locality requests (2) and (3) are equally incompatible with relativity makes  absolutely ununderstandable the position which  Bell has  taken concerning the dynamical reduction models. To understand this point we will pass now to analyze the recent debate on theories which overcome the measurement problem without conflicting   with  SR.

\section{ ``Exact" theories and their relativistic generalizations}

As is well known, the serious concern for the unsatisfactory status of standard QM with respect to the measurement problem has led various scientists to advance  proposals to overcome the difficulties. The one, besides Bohmian mechanics, which has been considered as fully satisfactory by Bell himself is the so called GRW-theory \cite{grw}. Its essential elements are well known: one considers nonlinear and stochastic modifications of  Schr\"{o}dinger's equation in such a way not to change all predictions of the standard theory concerning microscopic situations and leading to a consistent derivation of the process of reduction of the wave packet every time in which the standard theory would lead to superpositions of macroscopically different states. Bell has paid a lot of attention and has devoted various papers to this proposal.

To illustrate the concluding sentences of the last Section, we consider it important to call attention on the fact that Norsen's position puts (in a certain sense) into evidence what might look as a contradictory attitude of John Bell. In fact, on the one hand, he knew very well that ``collapse models"  implied a violation of \{O.I.\} but not of \{P.I.\}, on the other, he has clearly and repeatedly claimed that they paved the way  to the overcoming  of the conflict between the two basic pillars of modern theory, quantum mechanics and relativity. 

To support this view, we will recall that since his first presentation of the theory \cite{bellGRW}   and, subsequently, on various  occasions, he has paid a lot of attention to the potential conflict of the GRW-theory with SR. For example, in  \cite{schwinger} he claims: 
\begin{quote}
{\it ...when I saw this theory first, I thought that I could blow it out of the water, by showing that it was grossly in violation of Lorentz invariance.} [then he goes on describing the EPR like puzzle, and remarks] {\it ... so you see that what conventionally happens when we declare the firing of the counter to be a ``measurement" happens here just because the counters are large bodies.... Now this is very curious, not because of what happens on the right. When the spin-down state disappears on the left, the spin-up disappears on the right, and vice-versa. This is the ``spooky action at a distance" which Einstein saw in the ordinary formulation of quantum mechanics. That spooky action at a distance is not being removed by this reformulation, but emphasized. In the ordinary formulation you have a way out: maybe the wavefunction is not real - maybe there are things which are real which you never describe - maybe the things which are real are not behaving in a funny way. But in this new theory, there is nothing else but the wavefunction and the wavefunction is behaving in a funny way. It looks as if this could not possibly be Lorentz invariant, because of the long range instantaneity.}
\end{quote}

After having made this clear statement  he discusses the Lorentz invariance of the nonrelativistic description of dynamical reduction and he reproduces exactly the original argument that he had presented at the Imperial College Celebration. The proof that he gives, by resorting to the consideration of a two-times Schr\"{o}dinger's equation, amounts essentially to show that the theory does not violate  \{P.I.\}. Successively we have proved \cite{fleming1,fleming2} in general that the original GRW theory as well as the ``linear plus cooking" version of CSL (the  simplest continuous generalization of GRW) actually do not violate  \{P.I.\}. 

Since there is no doubt that the considered theories make predictions which agree to an extremely high degree of accuracy (practically they are identical) with the predictions of standard QM for the  EPR like situation we have discussed in this paper, they turn out to violate $\it \{Fact\}$, so that they must violate \{O.I.\}. This being the situation, how can we ignore that Bell himself has declared \cite{bellGRW}, with reference to the GRW theory:
\begin{quote}
{\it I am particularly struck by the fact that the model is as Lorentz invariant as it could be in the nonrelativistic version: It takes away the grounds of my fear that any exact formulation of quantum mechanics must conflict with fundamental Lorentz invariance}.
\end{quote}

This, in my opinion, illustrates  in a simple and lucid way that, for Bell, a theory violating \{O.I.\} can peacefully cohexist with SR, a point on which Norsen has been silent.

\section{Conclusions}
The analysis of the last part of this paper has called  the due attention, we hope, to the fact that when discussing the compatibility of a nonlocal theory with SR it is conceptually extremely relevant to analyze whether locality is violated because of a violation of \{P.I.\} or of \{O.I.\}. Moreover, it is important to take into account that today we can find in the literature interesting examples of relativistic dynamical reduction models. The possibility of working out such theories has attracted a lot of attention. 

Already the original (unsatifactory) field theoretical relativistic generalization of the GRW theory  presented in refs.\cite{pearle,ggp} has been proved \cite{ggp2} to fully satisfy relativistic requirements in the genuine sense. However, such a proposal met various difficulties related to the occurrence of untractable divergencies (which however are not related to its relativistic structure). Subsequently we have introduced \cite{ghirardi2000} a dynamical reduction toy model which, in spite of its oversimplified and unphysical structure, is manifestly relativistically invariant. In recent times  R. Tumulka has presented \cite{tumulka} a perfectly consistent and formally rigorous relativistic dynamical reduction model for many noninteracting fermions. D. Bedingham \cite{bedingham},  inspired by the model of ref.\cite{ghirardi2000},  has recently succeeded in  formulating  a well defined formal  theory inducing reductions which is compatible with SR.

In view of these facts we would like to conclude by remarking that:
\begin{itemize}
\item For testing the ``genuine" or ``instrumental" compatibility of a nonlocal theory with SR it is extremely important to investigate whether its nonlocal character derives from a violation of \{P.I.\} or of \{O.I.\},
\item Theories which violate \{P.I.\} admit at most relativistic generalizations which require the adoption of a (hidden) preferred reference frame. Accordingly they are only instrumentally relativistically invariant, but they can be physically fully satisfactory.
\item It turns out to be possible to exhibit genuinely relativistic nonlocal theories which violate only \{O.I.\}.
\end{itemize}

Our conclusions can be summarized in an extremely appropriate way by a recent statement made by R. Tumulka \cite{tumulka} when proposing his relativistic dynamical reduction model:
\begin{quote}
{\it A somewhat surprising feature of the present situation is that we seem to arrive at the following alternative: Bohmian mechanics shows that one can explain quantum mechanics, exactly and completely, if one is willing to pay with using a preferred slicing of space-time; our model suggests that one should be able to avoid a preferred slicing if one is willing to pay with a certain deviation from quantum mechanics.}
\end{quote}

We hope that the present analysis has to some extent  enriched the one presented by Norsen by stressing that, in spite of the inappropriate use that Jarrett makes of his {\it Locality} and {\it Completeness} requests, the identification that the fundamental request of factorizability is the logical conjunction of \{P.I.\} and \{O.I.\} remains an important tool for investigating the compatibility of quantum nonlocality with relativistic requirements.

\end{document}